\documentclass{aa}  
\usepackage{graphicx}
\usepackage{float}
\usepackage[colorlinks=true, linkcolor=blue, citecolor=black, urlcolor=blue, pdfborder={0 0 0}]{hyperref}\hypersetup{}
\usepackage{txfonts}
\usepackage[switch]{lineno}

\begin{document} 

\title{Fermi-LAT detections of the classical novae V1723 Sco and V6598 Sgr in a multi-wavelength context.}

\titlerunning{Fermi-LAT detections of novae V1723 Sco and V6598 Sgr.}

   \author{P. Fauverge\inst{1} \and 
            P. Jean\inst{2} \and
          K. Sokolovsky\inst{3}\and
           C.C. Cheung\inst{4}\and
          M. Lemoine-Goumard\inst{1}\and
          M.-H. Grondin \inst{1}\and
          L. Chomiuk \inst{5}\and
          A. Dickenson \inst{3}\and
          J. D. Linford\inst{6}\and
          K. Mukai \inst{7}\and
          J. L. Sokoloski\inst{8}
          }

   \institute{Univ.  Bordeaux, CNRS, LP2i Bordeaux, UMR 5797, F-33170 Gradignan, France\\
    \email{fauverge@lp2ib.in2p3.fr}
    \and
     IRAP, Université de Toulouse, CNRS, CNES, UPS, 9 avenue Colonel Roche, 31028 Toulouse, Cedex 4, France 
    \and
    Department of Astronomy, University of Illinois Urbana-Champaign, 1002 W. Green Street, Urbana, IL 61801, USA
    \and 
   Space Science Division, Naval Research Laboratory, Washington, DC 20375-5352, USA 
   \and
   Center for Data Intensive and Time Domain Astronomy, Department of Physics and Astronomy, Michigan State University, 567 Wilson Rd, East Lansing, MI 48824, USA 
   \and
   National Radio Astronomy Observatory, Domenici Science Operations Center, 1003 Lopezville Road, Socorro, NM 87801, USA
   \and
   CRESST and X-ray Astrophysics Laboratory, NASA/GSFC, Greenbelt, MD 20771, USA
   \and
   Department of Physics and Columbia Astrophysics Laboratory, Columbia University, New York, NY 10027, USA \\
             }

\abstract{
\textit{Context.} Numerous classical novae have been observed to emit $\gamma$-rays (E > 100 MeV) detected by the Fermi-LAT. The prevailing hypothesis attributes this emission to the interaction of accelerated particles within shocks in the nova ejecta. However, the lack of non-thermal X-ray detection coincident with the $\gamma$-rays remains a challenge to this theory. \

\textit{Aims.} We aim to constrain the $\gamma$-ray production mechanism by combining optical and X-ray data with a detailed analysis of the Fermi-LAT observations for two classical novae, V1723 Sco 2024 and V6598 Sgr 2023. \

\textit{Methods.} We performed similar analyses of the Fermi-LAT data for both novae to determine the duration, localization, and spectral properties of the $\gamma$-ray emission. These results were compared with optical data from the AAVSO database and X-ray observations from NuSTAR, available for V1723 Sco 2024 only, to infer the nature of the accelerated particles. Finally, we used a physical emission model to extract key parameters related to particle acceleration. \

\textit{Results.} V1723 Sco 2024 was found to be a very bright $\gamma$-ray source with an emission duration of 15 days allowing us to constrain the spectral index and the total energy of accelerated protons. Despite early NuSTAR observations, no non-thermal X-ray emission was detected simultaneously with the $\gamma$-rays. However, unexpected $\gamma$-ray and thermal hard X-ray emission were observed more than 40 days after the nova outburst, suggesting that particle acceleration can occur even several weeks post-eruption. V6598 Sgr 2023, on the other hand, was detected by the Fermi-LAT at a significance level of $4\sigma$ over just two days, one of the shortest $\gamma$-ray emission durations ever recorded, coinciding with a rapid decline in optical brightness. Finally, the high ratio of $\gamma$-ray to optical luminosities and $\gamma$-ray to X-ray luminosities for both novae, as well as the curvature of the $\gamma$-ray spectrum of V1723 Sco below 500 MeV, are all more consistent with the hadronic than the leptonic scenario for $\gamma$-ray generation in novae.}


   \maketitle

\section{Introduction}
Classical novae are thermonuclear eruptions that occur on the surface of a white dwarf (WD) that accretes matter from a main sequence star in a binary system. The thermonuclear eruption leads to an ejection of matter. This expansion is most often aspherical and can be approximated into two lobes  above and below the equatorial plane of the binary system \citep{metzger_gamma-ray_2015}. P-Cygni profiles of optical spectral lines in novae indicate expanding material reaching velocities of thousands of km s$^{-1}$ over several days. In addition, the ejecta can be shocked by particles from the radiation-driven wind from the very hot WD, which can lead to the creation of several internal shocks in the system \citep{chomiuk_new_2021}. These shocks appeared to be very efficient at accelerating particles via diffusive shock acceleration (DSA; \citet{DSA}) up to several tens of GeV. These particles can then interact with the ambient medium and finally emit photons from the radio \citep{radionovae} to $\gamma$-ray band \citep{novae_fermi}, covering the range of X-ray \citep{2020MNRAS.497.2569S} and optical wavelengths. This emission is produced alongside the ordinary thermal emission of the hot nova shell. Up to date, 24 classical novae have been detected by the Fermi-Large Area Telescope (LAT)\footnote{see \href{https://asd.gsfc.nasa.gov/Koji.Mukai/novae/novae.html}{Koji's List of Recent Galactic Novae}} \citep{novae_fermi} and the $\gamma$-ray peaks were typically close in time to the optical ones. Additionally, in one of them, the time correlation between several optical and $\gamma$-ray peaks allowed \citet{2020NatAs...4..776A} to conclude that both emissions come from the same region and confirm that the shocks could be radiative.

In the case of the X-rays, the emission can be divided into three parts. The first one is the "fireball phase," a bright soft thermal flash that occurs before the outburst of the nova \citep{fireball}. Then the X-ray emission is dominated by the shock-powered thermal emission at energies$\sim$ 1-10 keV \citep{v3890sgr_chandra}. Finally, the so-called Super Soft Source (<0.5 keV; SSS) phase starts when the layer of matter ejected is diluted and reveals the hot WD surface \citep{SSS_swift}. A non-thermal X-ray emission is also expected, by extrapolation down to keV, but has never been observed to date \citep{2020MNRAS.497.2569S}. The two most recent novae detected by Fermi-LAT at the time of writing represent the variety of classical $\gamma$-ray novae, with one being very bright and long-lasting, while the other is among the shortest ever detected.

The nova V1723 Sco (also known as Nova Sco 2024 or PNV J17261813-3809354) was discovered in optical by A. Pearce on 2024 Feb 08.827 with a magnitude of 7.8 \citep{cbet_v1723sco}. Approximately 1 day later, \citet{2024ATel16439....1C} announced a detection in $\gamma$-rays with the Fermi-LAT, during a bin of 6 hours between Feb 9.5 and Feb 9.75 with a significance of 4.2$\sigma$ assuming a power-law spectral model. The analysis of a 1-day bin finally showed a very significant (> 5$\sigma$) signal between Feb 11.0 and Feb 12.0 \citep{2024ATel16441....1C}. The optical transient was classified as a nova on February 10.4 UT using the SOAR telescope, based on the detection of absorption lines and P Cygni profiles, as reported by \citet{2024ATel16440....1A}. NuSTAR \citep{nustar} and Swift \citep{swift} observations confirmed the thickness of the eruption in the X-ray band during the day following the eruption with non-detection until Feb 22.268 ($\sim 13.5$ days after the optical discovery) and a peak in the X-ray data 33 days after the eruption \citep{2024ATel16444....1S,2024ATel16484....1S}. Finally, V1723 Sco was also monitored by the VLA radio telescopes starting from Feb 17 \citep{2024ATel16492....1M}.

The nova V6598 Sgr (also known as Nova Sgr 2023 N.3 or TCP J17525020-2024150), was discovered by A. Pearce on 2023 Jul 15.459 in optical \citep{cbet_v6598sgr} and was then classified as a classical nova on Jul 15.95 \citep{2023ATel16135....1G}. The binary system that hosts this nova is known as the hard X-ray source IGR J17528-2022 \citep[J17528 in the following,][]{cat_integral} discovered with INTEGRAL \citep{integral_paper}. \citet{chandra_nustar} performed a Chandra and NuSTAR analysis on J17528 and classified it as a cataclysmic variable (CV). The hard X-ray component could even point toward a magnetized CV (mCV) but the non-detection of either spin or orbital period cannot confirm its type (polar or intermediate polar). It was the first time that a classical nova was characterized in X-rays so precisely before the eruption. \citet{chandra_nustar} also derived the pre-eruption optical spectrum of J17528 characterized by a strong $H_\alpha$ as well as weak He I emission. They also used Chandra/NuSTAR X-ray spectra to derive the large absorbing column of $N_H \sim 3.2 \times 10^{22}$\,cm$^{-2}$. The optical spectroscopy by \citet{2023ATel16135....1G} during the nova eruption revealed the interstellar Na I D absorption lines with their equivalent widths corresponding to $N_H \sim 4 \times 10^{21}$\,cm$^{-2}$ and implying that much of the pre-eruption X-ray absorption was intrinsic to the binary. This nova was also announced as a 4$\sigma$ detection by the Fermi-LAT \citep{2023ATel16151....1J} during the two days interval Jul 15-17 (assuming a power-law spectrum). In the X-ray band, Swift-XRT did not detect any source at the nova's position shortly after the outburst, even though the host system had previously emitted X-rays, which further confirms that X-rays were absorbed in the days following the eruption. A source only reappeared 17 days later, with a brightness approximately four times lower than the pre-outburst flux \citep{2023ATel16172....1N}. Finally, the ASKAP telescopes detected 888 MHz radio emission 93 days after the eruption \citep{2023ATel16383....1D} with a flux density of 4.5 $\pm$ 0.3 mJy. The previous observation, on 2023 Oct 02, 79 days after the outburst, did not yield a detection with a 5$\sigma$ upper limit of 1.3 mJy. 

In this paper, the analysis of the Fermi-LAT data of the two novae will be presented in Section 2, as well as the observations in the optical from the AAVSO database\footnote{Data for both novae are available in: \url{https://www.aavso.org/data-access}} \citep{AAVSODATA} and the X-ray band using NuSTAR data. Multi-wavelength results and their constraints on the physical parameters of the sources will be discussed in Section 3, and the conclusion will be presented in Section 4.

\section{Observations and analyses}
\subsection{Fermi-LAT data analysis procedure}
\begin{figure*}
\centering
    
    \includegraphics[width=17cm]{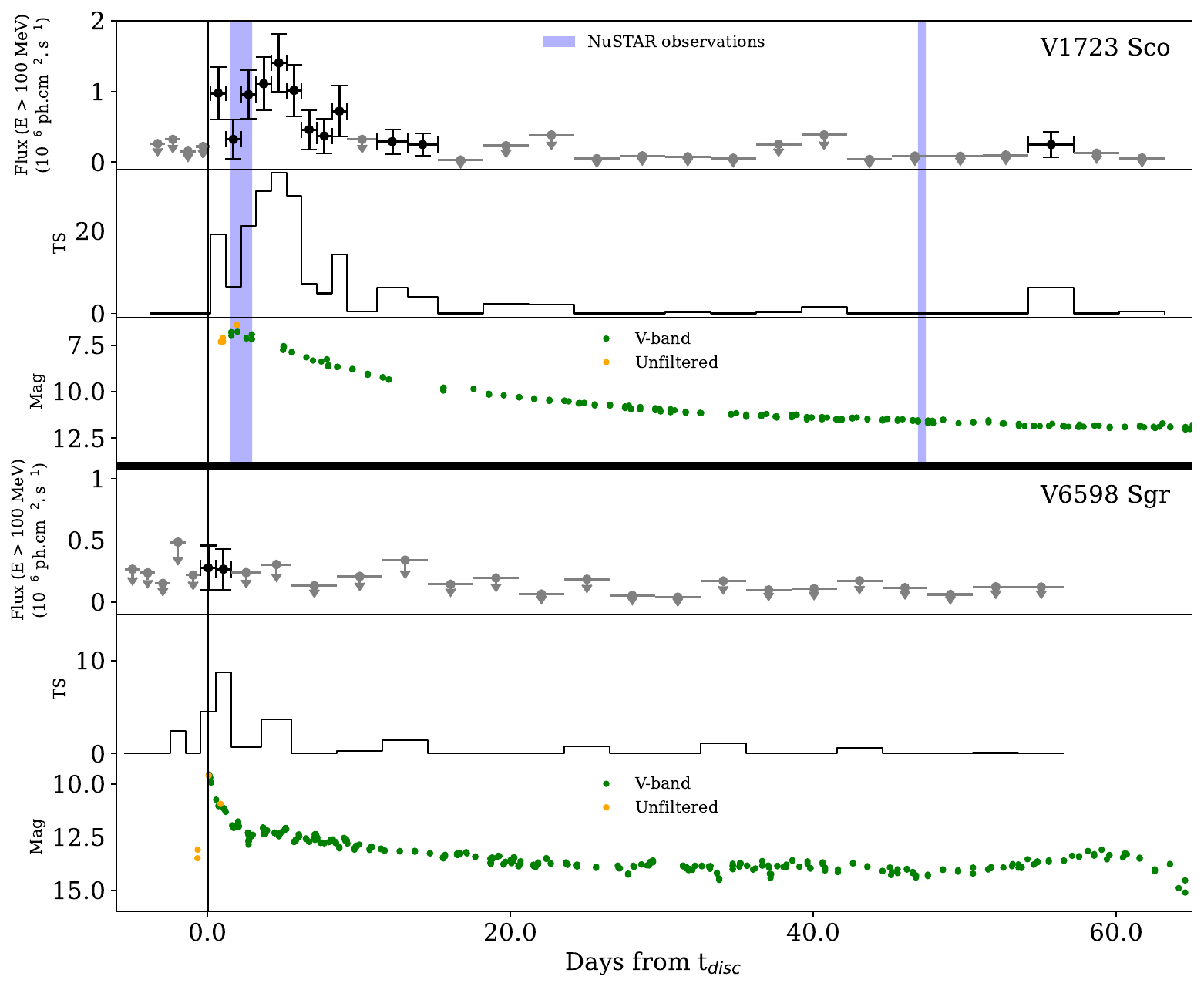}
    
    \caption{\label{fig:lightcurves} {$\gamma$-ray and optical light curves of V1723 Sco (top) and V6598 Sgr (bottom) calculated from the Fermi-LAT data (black and gray) and taken from the AAVSO database in the V-band (green) and unfiltered (orange) respectively. Flux estimates with 1$\sigma$ statistical uncertainties only are plotted when TS and $N_{pred}$ > 4  and 95 $\%$ upper limits otherwise. The blue windows indicate the time periods of the NuSTAR observations. $t_{disc}$ corresponds to the optical discovery of both novae (Table \ref{tab:params}}).}
\end{figure*}

For the Fermi-LAT \citep{fermilat} analyses of the two novae, we used Pass 8 data \citep{PASS82013,pass82018}, associated with the \texttt{P8R3\_SOURCE\_V3} instrument response function, within a 15° region centered on a point slightly offset from the optical nova localization (Table \ref{tab:params}) to avoid being at the intersection between pixels. We selected all photons between 100 MeV and 300 GeV with a maximum zenith angle of 90°, and binned the data with spatial bins of 0.1° and eight logarithmically spaced energy bins per decade. The spatial and spectral models of our region were built using the 4FGL catalog based on 14 years of LAT data \citep{4fgl_dr3,4fgl_dr4} selecting all the sources within 20° of the region of interest (ROI) center. The galactic diffuse component was modeled by the \texttt{gll\_iem\_v07.fits} template, and the isotropic component (which includes isotropic background and instrumental noise), was modeled with \texttt{iso\_P8R3\_SOURCE\_V3\_v1.txt}\footnote{Both files are available in: \url{https://fermi.gsfc.nasa.gov/ssc/data/access/lat/BackgroundModels.html}}. Finally, all analyses were performed using version 1.2.0 of \texttt{fermipy} \citep{fermipy}.

In both analyses, the method is the following. First, the model of the ROI is optimized on a one-year period ending more than one week before the discovery of the novae in optical ($t_{disc}$). Then, we compute the light curve of a point source at the nova position and choose the optimal time interval of the $\gamma$-ray emission. Finally, during this optimal time interval, we re-localize the $\gamma$-ray emission, and find the best spectral model. All the parameters are summarized in Table \ref{tab:params} for the two novae.

\paragraph{1-year period analysis}
Most of the sources in the ROI are not bright enough to be well modeled during a time period of only 2 weeks \citep[classical for nova $\gamma$-ray emission][]{novae_fermi}. To refine our starting model, an analysis of a 1-year period before the outburst (see Table \ref{tab:params}) was performed \citep{rsoph_fermi}. The two diffuse components and the brightest sources were fitted. As a test, a point source at the position of the nova, assuming a power-law spectral model, was added and fitted subsequently, but no significant signal was found. This model (which will be referred to as $M_0$, without any source at the nova position) is taken for the following analyses as a reference because none of the brightest sources are expected to be variable during the short time interval of the outburst.

\paragraph{Time duration determination}
Novae are known as transient sources, so, to perform a spectral analysis, a time period has to be chosen. In order to estimate it, a point source with a power-law spectrum at the optical nova position is added to $M_0$ and a light curve (evolution of the flux with respect to time) is calculated over 70 days. The light curve was calculated with 1-day time bins. In each of these bins, the amplitude is free and the index is fixed to 2. Flux points are calculated when the test statistic TS \citep{ref_ts} and the number of predicted photons from the source, $N_{pred}$, are higher than 4 ( see Figure \ref{fig:lightcurves}) which is equivalent to a 2$\sigma$ detection; 95$\%$ upper limits are computed otherwise. In a second step, 1-day bins are allowed to merge into 2 or 3-day bins to look for a potential 2$\sigma$ detection (as can be seen between day 11 and 15 in Figure \ref{fig:lightcurves} for V1723 Sco). We define the duration of the outburst as the time interval between the start of the first and the end of the last bin with 2$\sigma$ detection, allowing for a maximum time gap with upper limits of two days.

\begin{figure*}[!htbp]
\sidecaption
    \includegraphics[width=6cm]{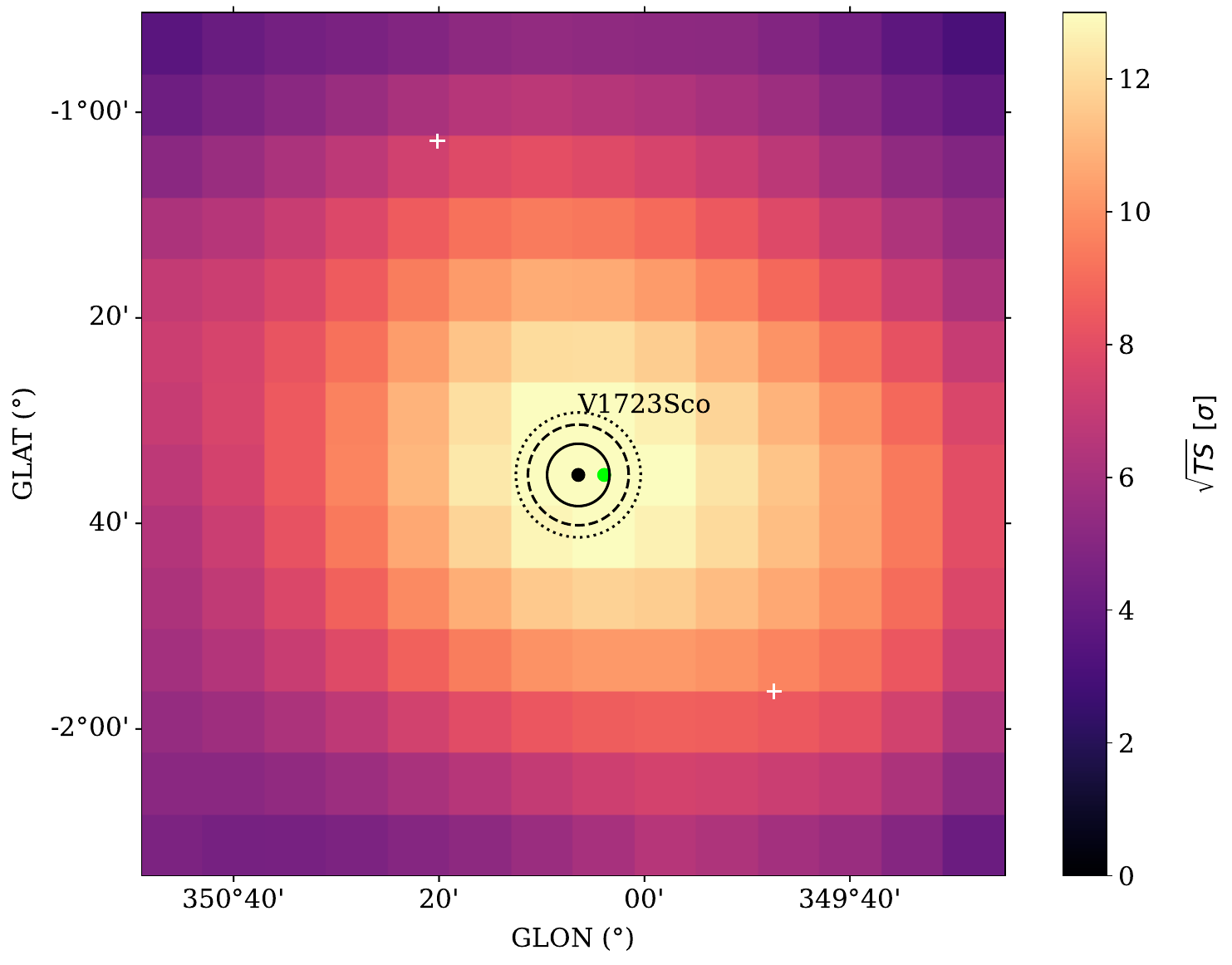}
    \includegraphics[width=6cm]{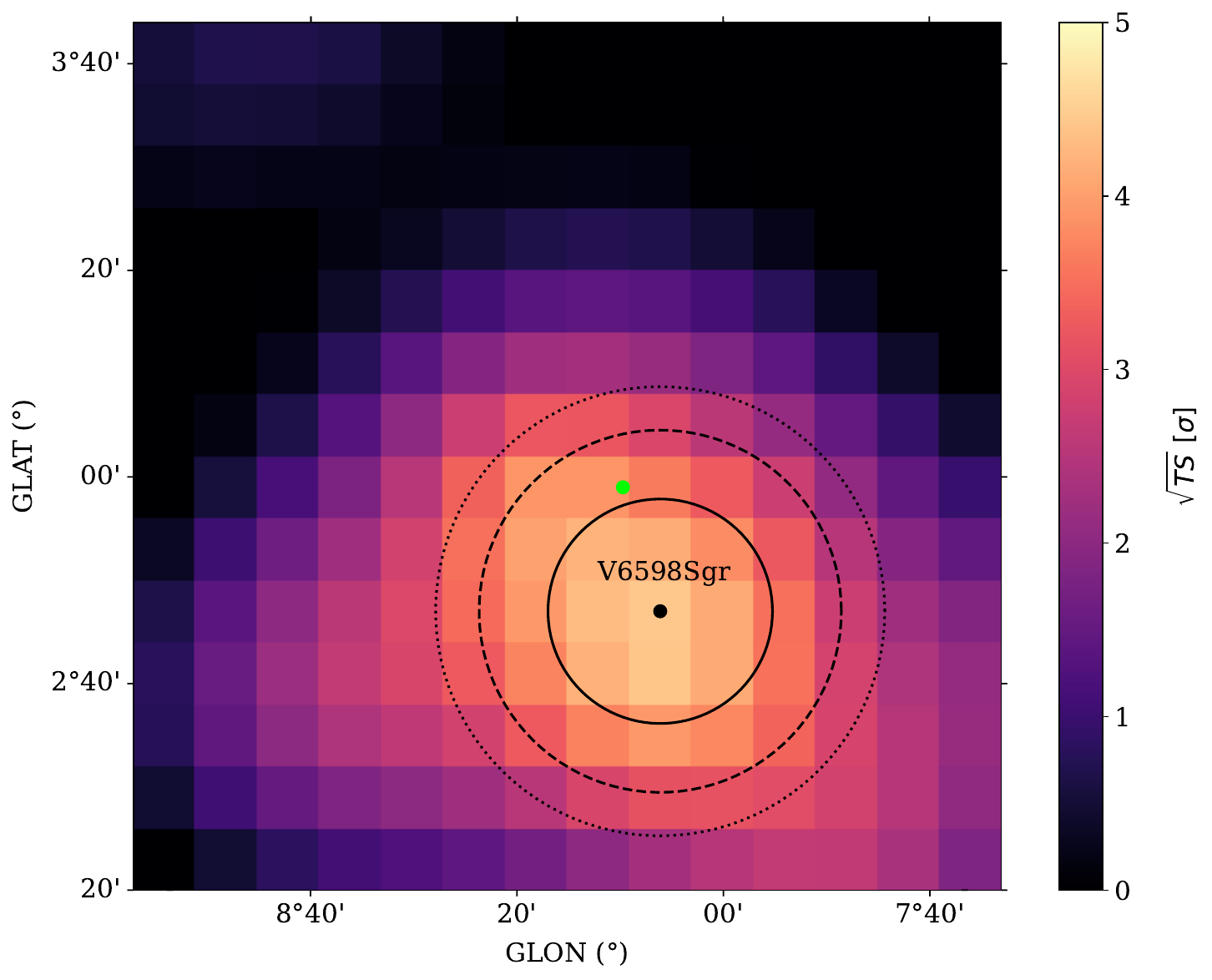}
        \caption{\label{fig:tsmaps} {TS maps of V1723 Sco (left) and V6598 Sgr (right) during the optimal time period (see text). The value in TS is estimated in each pixel by fitting a point source with a power-law model (index fixed to 2). The green point is the optical nova position, the black one is the position found by \texttt{gta.localize} with the 68$\%$, 95$\%$ and 99$\%$ containment radius. The white crosses are known $\gamma$-ray sources from the 4FGL catalog.}}
\end{figure*}

\paragraph{Localization and spectral analysis}
The position of $\gamma$-ray and optical emission are not expected to be different, however, for consistency, we verified if both positions aligned within the Fermi-LAT angular uncertainty of the source\footnote{The Fermi-LAT performance is summarized in: \url{https://www.slac.stanford.edu/exp/glast/groups/canda/lat_Performance.htm}}.
For each of the spectra presented below, the tool \texttt{gta.localize} is applied to the point source initialized at the optical nova position. This tool localizes the peak in the likelihood spatial distribution of the region and estimates the 68, 95, 99 $\%$ containment radii (Figure \ref{fig:tsmaps}).

\begin{figure*}
    \centering
    \includegraphics[width=8.5cm]{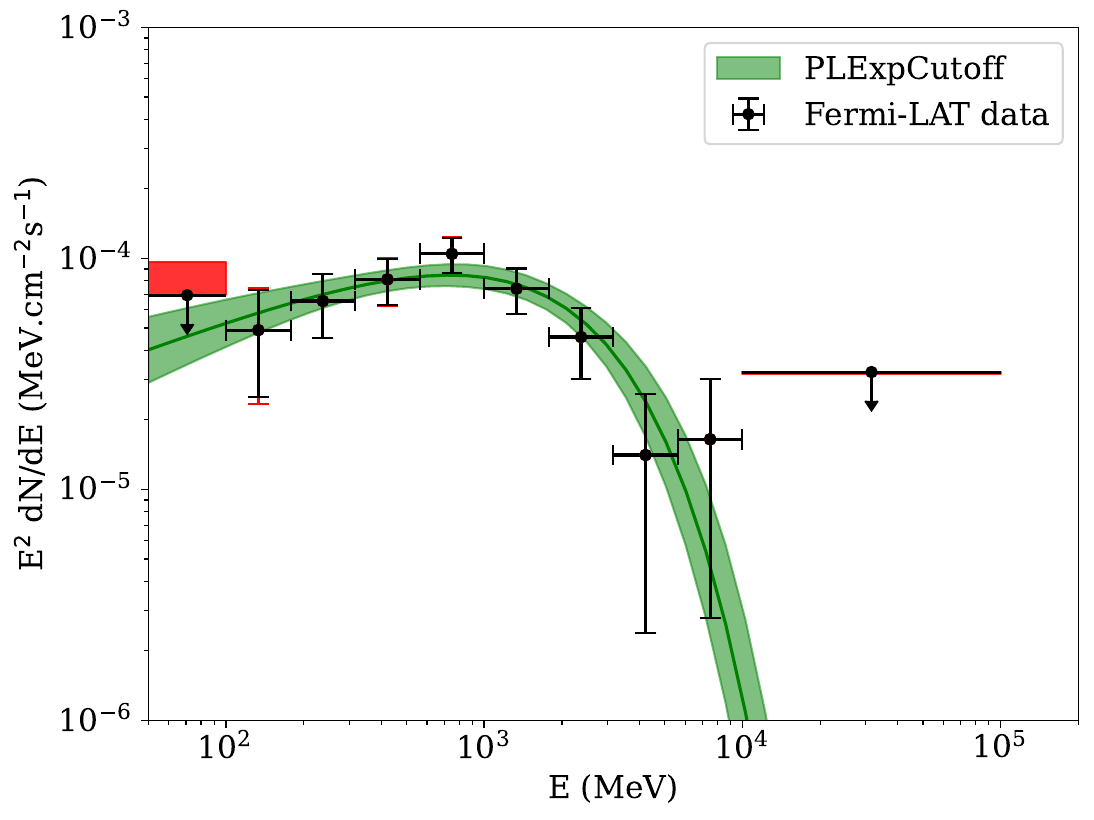}
    \includegraphics[width=8.5cm]{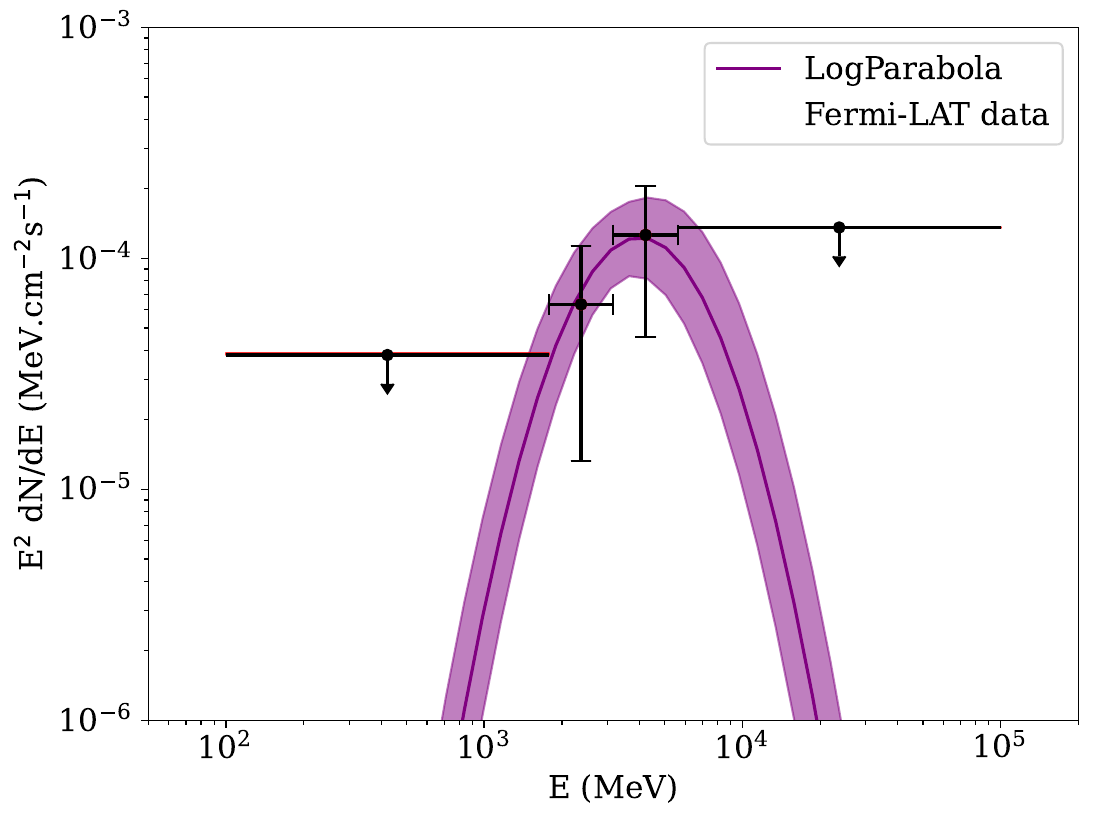}    
    \caption{\label{fig:SEDs} {Spectra of V1723 Sco (left) and V6598 Sgr (right) during the optimal time period (see text). Only the best spectral model and its corresponding uncertainty was represented. Flux points with 1$\sigma$ statistical uncertainties (black) and the quadratic sum of statistical and systematic uncertainties (red) are plotted when TS and $N_{pred}$ > 2  and 95 $\%$ upper limits otherwise. Energy bins from the analysis are merged only for the calculation of the flux points.}}
\end{figure*}

Three spectral models were tested for a point source at the optical position: a power-law (PL), a logarithmic parabola (LogP), and a power-law with an exponential cutoff (PLExpCutoff)\footnote{All the models implemented in the \texttt{fermitools} are defined in: \url{https://fermi.gsfc.nasa.gov/ssc/data/analysis/scitools/source_models.html}}. The amplitude and the index were free for all the models, in addition to the cutoff energy and $\beta$ for the PLExpCutoff and the LogP respectively. For the three models, the fit was done with only the nova's parameters and the amplitude of the isotropic diffuse component free to vary, while all other parameters are fixed to their 1-year period values from $M_0$. Tests performed by fixing the isotropic diffuse component do not change the results of the nova parameters. The best fit model and associated parameters are presented in Table \ref{tab:params} (bottom). The spectral energy distribution (SED) is calculated with a power-law model in each energy bin, with only the amplitude of the nova emission free to evolve and the index fixed to 2 (see Figure \ref{fig:SEDs}). The systematic uncertainties are calculated by replacing the galactic diffuse model with eight alternative ones developed by the Fermi collaboration \citep{snrcat}. The systematic uncertainties due to uncertainties on the effective area were also calculated, but were found to be negligible.

\begin{table}[ht]
    \caption{\label{tab:params}{Parameters of the Fermi-LAT analysis for the two novae.}}
    \resizebox{\columnwidth}{!}{
    \begin{tabular}{lcc}
        \hline \hline
        \noalign{\smallskip}
        \multicolumn{1}{l}{} & V1723 Sco & V6598 Sgr \\ 
        \noalign{\smallskip}
        \hline
        \noalign{\smallskip}
        \multicolumn{1}{l}{Optical nova position (°)} & 
        \begin{tabular}[c]{@{}c@{}}RA : 261.5755 \\ Dec :  -38.1598 \\ glon : 350.0654\\ glat : -1.5882
        \end{tabular} & 
        \begin{tabular}[c]{@{}c@{}}RA : 268.2054  \\ Dec : -20.4042 \\glon : 8.1621\\ glat : 2.9835
        \end{tabular} \\ 
        \noalign{\smallskip}
        \hline
        \noalign{\smallskip}
        \multicolumn{3}{c}{1-year analysis} \\ 
        \noalign{\smallskip}
        \hline
        \noalign{\smallskip}
        \multicolumn{1}{l}{1-year period} & 
            \begin{tabular}[c]{@{}c@{}}2023 Feb 01\\ 2024 Feb 01
            \end{tabular} & 
            \begin{tabular}[c]{@{}c@{}}2022 Jul 01\\ 2023 Jul 01
            \end{tabular} \\ 
        \noalign{\smallskip}
        \hline
        \noalign{\smallskip}
        \multicolumn{3}{c}{Time duration determination}\\ 
        \noalign{\smallskip}
        \hline
        \noalign{\smallskip}
        \multicolumn{1}{l}{Optical discovery ($t_{disc}$)}& 2024 Feb 08.827 & 2023 Jul 15.459 \\
        \multicolumn{1}{l}{$t_0$}& 2024 Feb 09.0 & 2023 Jul 15.0  \\
        \multicolumn{1}{l}{Optimal time period} & $t_0$ --- $t_0$+15 days & $t_0$ --- $t_0$ + 2 days \\ 
        \noalign{\smallskip}
        \hline
        \noalign{\smallskip}
        \multicolumn{3}{c}{Localization and spectral analysis}\\ 
        \noalign{\smallskip}
        \hline
        \noalign{\smallskip}
        \multicolumn{1}{l}{$\Delta$TS$_{\text{reloc}}$}& 0.3 & 5.8  \\
        \multicolumn{1}{l}{Offset (°)}&  0.03 & 0.22 \\
        \multicolumn{1}{l}{95 $\%$ containment radius (°)} &  0.08 &  0.29 \\        
        \multicolumn{1}{l}{Optimized position (°)} & -  & 
            \begin{tabular}[c]{@{}c@{}}glon : 8.102 \\ glat : 2.784
            \end{tabular}\\ 
        \multicolumn{1}{l}{Best fit model}& PLExpCutoff & LogParabola \\
        \multicolumn{1}{l}{TS}& 216 & 24.5 \\
        \multicolumn{1}{l}{\begin{tabular}{@{}l@{}}Flux (E \textgreater 100 MeV) \\ (10$^{-6}$ ph cm$^{-2}$ s$^{-1}$)\end{tabular}} & 1.1$^{+0.3}_{-0.2}$ & 0.05 $^{+0.03}_{-0.02}$ \\
        \multicolumn{1}{l}{$\Gamma$} & 1.5 $\pm$ 0.2 & - \\
        \multicolumn{1}{l}{E$_{Cutoff}$ (GeV)} & 1.6 $\pm$ 0.6 & - \\
        \multicolumn{1}{l}{$\alpha$} & - & -3.3 $\pm$ 1.8 \\
        \multicolumn{1}{l}{$\beta$}  & - & 1.9 $\pm$ 0.6 \\ 
        \noalign{\smallskip}
        \hline
    \end{tabular}
    }
    \tablefoot{The optical novae positions are from \citet{cbet_v1723sco} and \citet{cbet_v6598sgr} for V1723 Sco and V6598 Sgr respectively. The uncertainties on spectral parameters are 1$\sigma$ statistical uncertainties. The spectral parameters presented are those of the best fit model for both novae (see text).}
\end{table}

\subsubsection{Nova V1723 Sco 2024}

\begin{figure}[h]
        \includegraphics[width=1.\columnwidth]{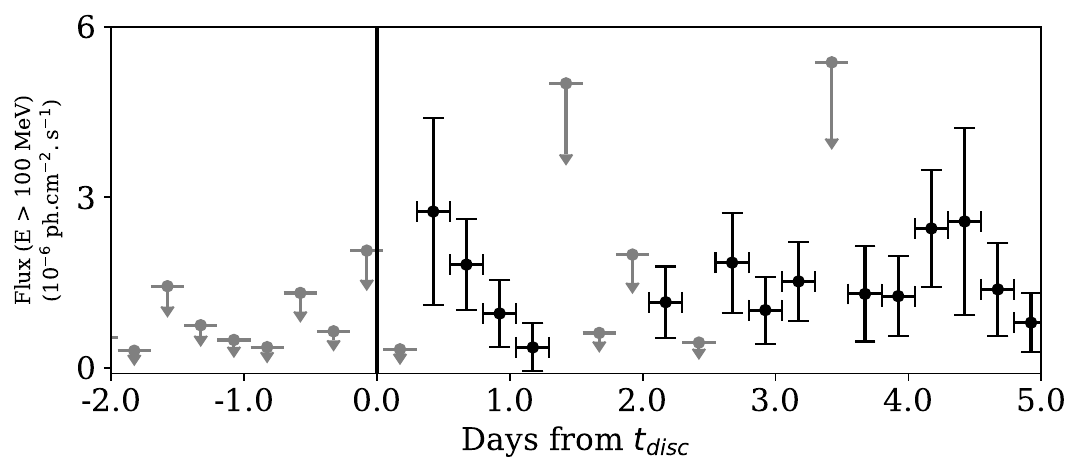}
    \caption{\label{fig:lightcurve_6hb} {Zoom into the Fermi-LAT light curve of V1723 Sco (Figure \ref{fig:lightcurves}) recalculated with bins of 6 hours length. See Figure \ref{fig:lightcurves} for the legend.}}
\end{figure}

The 1-year analysis before the outburst shows no emission from the binary system before the discovery of the nova. Thus, all the $\gamma$-ray emission from the source comes from the nova outburst and lasted approximately 15 days (Figure \ref{fig:lightcurves}). The first 9 bins have a duration of 1 day, followed by three 2-days long bins. To have a better precision on the start of the $\gamma$-ray emission, we repeat the light curve analysis with 6-hours bins (see Figure \ref{fig:lightcurve_6hb}). 

The 6-hour bins analysis reveals that the $\gamma$-ray emission started 6 hours after the optical discovery. Finally, a light curve for $t>t_{disc}+15 $ days is calculated for 3-days bin. Interestingly, one of these 3-day bins, centered at day 55, presents a TS and $N_{pred}$ >4. The characteristics of this time period will be discussed later.

The localization of the source in $\gamma$-rays confirms that $\gamma$-ray position is compatible, within the statistical uncertainties, with the optical nova localization, as seen in Figure \ref{fig:tsmaps}. This optical position is therefore used in the following for the spectral analysis.

In the spectral energy distribution presented in Figure \ref{fig:SEDs}, one can see the importance of systematic uncertainties at $E < 200$ MeV due to the uncertainties of the emission model from the galactic plane. However, because this nova is exceptionally bright in $\gamma$-rays, we expanded the energy range down to 50 MeV. Unfortunately, no significant detection was found in this energy bin and the systematic uncertainties due to uncertainties in the galactic diffuse emission model do not allow us to further constrain the spectral shape at low energy with this additional upper limit.
The nova outburst is significantly detected up to 10 GeV and is well modeled by a PLExpCutoff, which is significantly better (4.8$\sigma$) than the power-law model. The parameters of the best-fit model ($\Gamma$ and $E_{\text{Cutoff}}$; see Table \ref{tab:params}) are similar to those found with other novae detected by the LAT \citep{novae_fermi}.

A dedicated analysis of the short emission detected around $t_{{disc}} + 55$ days reveals a $\gamma$-ray excess slightly offset from the optical position, though still within the 95\% containment radius. The local TS is 11.7, and the short emission is well modeled by a power law with an index $\Gamma = 2.1 \pm 0.2$. No other variable source was found with TS $> 1$ within $9^\circ$ of the nova position. With four degrees of freedom, this corresponds to an excess of $2.3\sigma$ using the Wilks' theorem \citep{Wilks1938}. However, one must account for the number of trials required to observe such an excess. After the end of the Fermi detection, we tested 16 three-day time windows and selected one of them. Consequently, considering these 16 trials, the significance reduces to $1.0\sigma$. This level is too low to claim any detection, but the possible $\gamma$-ray excess should be considered together with the late-time NuSTAR detections of this source (see §~\ref{X-ray data}), which could provide interesting evidence for late-time particle acceleration in novae.

\subsubsection{Nova V6598 Sgr 2023}
In the case of V6598 Sgr, a source with a TS = 4.3 was observed during the one-year period before the outburst (equivalent to a 1.57$\sigma$ detection for a power-law with two degrees of freedom). We considered that this would not influence our model $M_0$ and proceeded as we did for V1723 Sco for the rest of the analysis. The time-duration analysis showed that $\gamma$-ray emission occurred simultaneously with the optical discovery and lasted only two days. The time bins were aligned with the first Fermi-LAT analysis performed by \citet{2023ATel16151....1J} to avoid having to account for any trials. The light curve in Figure~\ref{fig:lightcurves} confirms the two-day detection previously reported.

However, the localization of the source leads to an offset between the position of the peak in $\gamma$-ray data and the optical nova position of 0.22°,with an improvement of 2$\sigma$ ($\Delta$TS = 5.8 for 2 degrees of freedom more). The optical position is within the 95 $\%$ containment radius of 0.29°, and the offset might be due to the large point spread function (PSF) of Fermi at low energy (a 68$\%$ containment radius of 0.3° at 3 GeV) and should be interpreted as a statistical effect. Using this new position, the best fit model is a LogParabola (with $\Delta$ TS  = 4.4 with respect to the power-law model) and led to a TS of the source of TS = 24.5. This value is equivalent to a 3.8$\sigma$ detection, which is in agreement with the previous analysis done by \citet{2023ATel16151....1J} with a simple power-law in the same time interval. As an additional check, an analysis of 15 years of Fermi data prior to the outburst and two years after it, using two-day bins, leads to a probability of finding a bin with this TS coincident with the nova outburst by chance conservatively estimated at < 0.1$\%$. It is remarkable that no significant emission was found for this nova at energies below 1 GeV and it will be discussed in §\ref{chap.magnetic_field}.

\subsubsection{Placing V1723 Sco and V6598 Sgr in the context of $\gamma$-ray novae}
\begin{figure}[h]
        \includegraphics[width=1.\columnwidth]{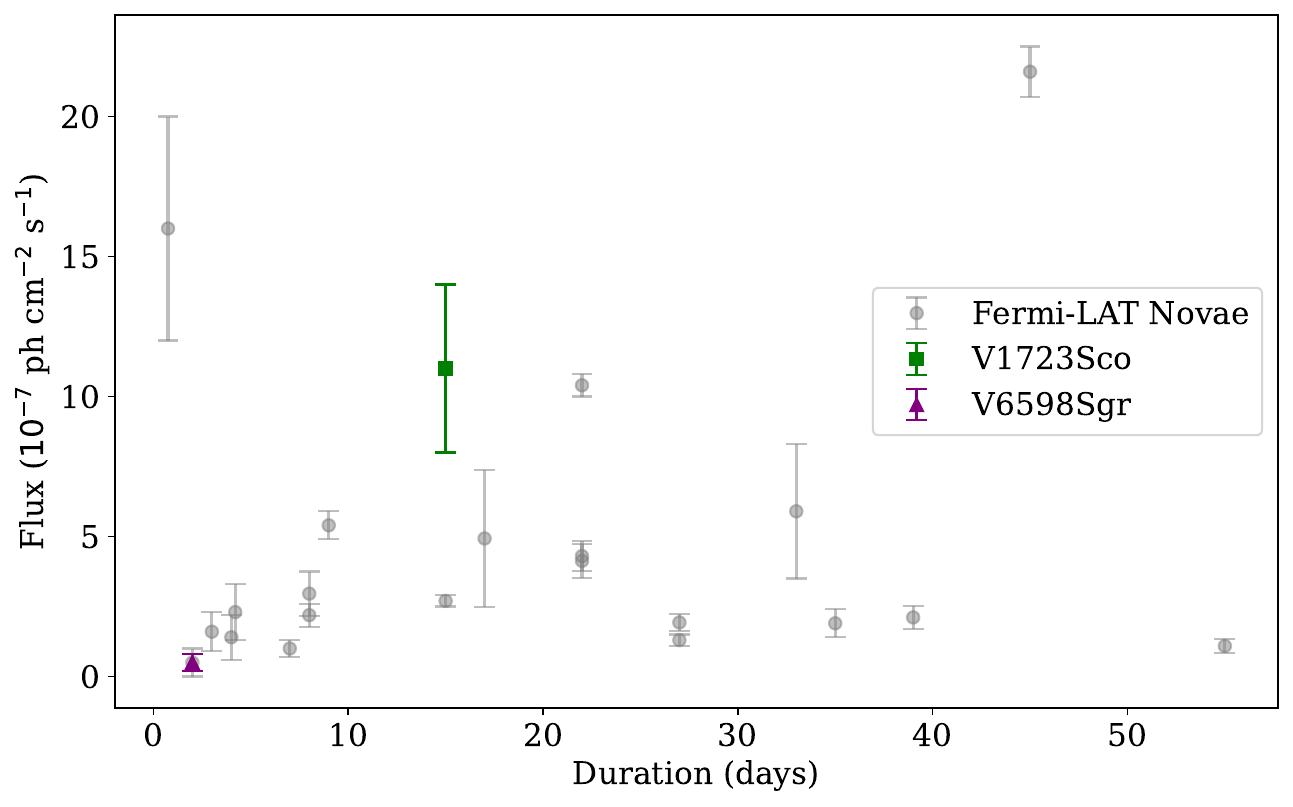}
    \caption{\label{fig:fluxvsduration} {V1723 Sco and V6598 Sgr in the flux-duration parameter space. We used average fluxes and duration measured by \citet{2018A&A...609A.120F}, \citet{novae_fermi}, \citet{2016ApJ...826..142C}, \citet{V407Lup..Gordon}, \citet{2019ApJ...872...86N}, \citet{lietal2017}, \citet{2020ApJ...905..114L}, \citet{2020NatAs...4..776A}, \citet{2022ApJ...940..141A}, \citet{2019ATel13116....1L}, \citet{2022MNRAS.514.2239S}, \citet{2021ATel14658....1B}, \citet{Her2021...Sokolovsky} and \citet{rsoph_fermi} for the Fermi-LAT novae points and this work for V1723 Sco and V6598 Sgr. The fastest and brightest nova is V1674 Her, the longest-lasting is V5668 Sgr, and the brightest and long-duration nova is RS Ophiuchi.}}
\end{figure}

In the context of the broader Fermi-LAT nova population, V1723 Sco and V6598 Sgr occupy distinct positions in the flux–duration parameter space (see Figure \ref{fig:fluxvsduration}). V1723 Sco shows a relatively high gamma-ray flux and an intermediate emission duration, placing it among the brighter and longer-lasting events compared to most other novae in the sample. In contrast, V6598 Sgr exhibits a much lower flux and very short duration, situating it at the faint, fast-evolving end of the distribution. These positions highlight the diversity of nova gamma-ray properties. The average flux was given instead of luminosities because of the large uncertainties in distance estimates for novae.

\subsection{Optical data}
For both novae, optical light curves were taken from the AAVSO database and are presented in Figure \ref{fig:lightcurves}. For V1723 Sco, the optical emission decreased monotonically, with no distinct features corresponding to the late $\gamma$-ray emission at day 55. The time to decline by two magnitudes $t_2$ is approximately 8 days.

In the case of V6598 Sgr, the optical luminosity declined rapidly. The time to fade by two magnitudes from peak brightness, $t_2$, is estimated to be approximately 2 days, making V6598 Sgr one of the fastest novae ever observed in the optical \citep[e.g.,][]{Her2021...Sokolovsky}. This is consistent with the very brief $\gamma$-ray emission detected, and supports the idea that both the optical and $\gamma$-ray signals originate from the same physical process, shocks \citep{metzger_gamma-ray_2015}. \citet{2020NatAs...4..776A} showed that in the case of V906 Car, the delays between the features appearing in the $\gamma$-ray and optical light curves were compatible with zero at the 2$\sigma$ level, further reinforcing this scenario.

\subsubsection{Distance estimate}\label{distance}
Using optical data, one can estimate the distances to the novae. In the following, calculations are detailed for the case of V1723 Sco while only their results are presented for V6598 Sgr. 

The method is based on the estimation of the absolute magnitude $M_{V,peak}$ with the Maximum Magnitude vs. Rate of Decline (MMRD) method: see Section 7.2 of \citet{schaefer_comprehensive_2022}. The empirical relation is 
\begin{equation}
M_{V,peak} = -7.6 +1.5 \ \text{log}\left(\frac{t_3}{30 \text{ days}}\right) \ \pm \ 1.3,
\end{equation}
with $t_3$ the time in days to decline by 3 magnitudes. With the light curve from AAVSO (Figure \ref{fig:lightcurves}), one can estimate $t_3 \sim 10$ days. The interstellar extinction has to be taken into account and can be estimated using \citet{gaver_relation_2009} with a column density $N_H$ at a value of $N_H \sim 8.5\times 10^{21}$ cm$^{-2}$ using \citet{2024ATel16444....1S}. This leads to $A_V = 3.8 \ \pm \ 0.16 $ mag. With a maximum apparent magnitude of 6.77 \citep{2024ATel16454....1H}, the distance from the nova V1723 Sco can be estimated to $D = 1.9\ \pm \ 1.1 $ kpc. This distance is in agreement with the value (1.7 $\pm$ 1.1) kpc obtained by assuming that the intrinsic color of the nova is of -0.03, two magnitudes below peak \citep{2025MNRAS.538.2339C} and a reddening between 1.1 and 1.5 \citep{2024ATel16492....1M}.

For nova V6598 Sgr, \citet{2023ATel16143....1D} estimated $A_V$ = 5.6 $\pm$ 1.0 mag and using the optical light curve from AAVSO (Figure \ref{fig:lightcurves}) a value of $t_3 \sim 3 $ days is estimated. That leads to a distance $D\sim5.8 \ \pm \ 3.7 $ kpc. With such a large uncertainty, we can not exclude the possibility that V6598 Sgr belongs to the galactic bulge ($d \sim 8 $ kpc). The MMRD method is an empirical relation that has shown its limitations \citep{2018MNRAS.481.3033S} and, in the absence of independent confirmation for this nova, it should be used with caution, especially given that the properties of the host system are unusual and extreme for a CV, assuming the large distance estimate is correct.

\subsubsection{Flux Ratio $F_{\gamma}/F_{\text{opt}}$ \label{chap.lgamma/lopt}}
\begin{figure}[h]
     \includegraphics[width=1\columnwidth]{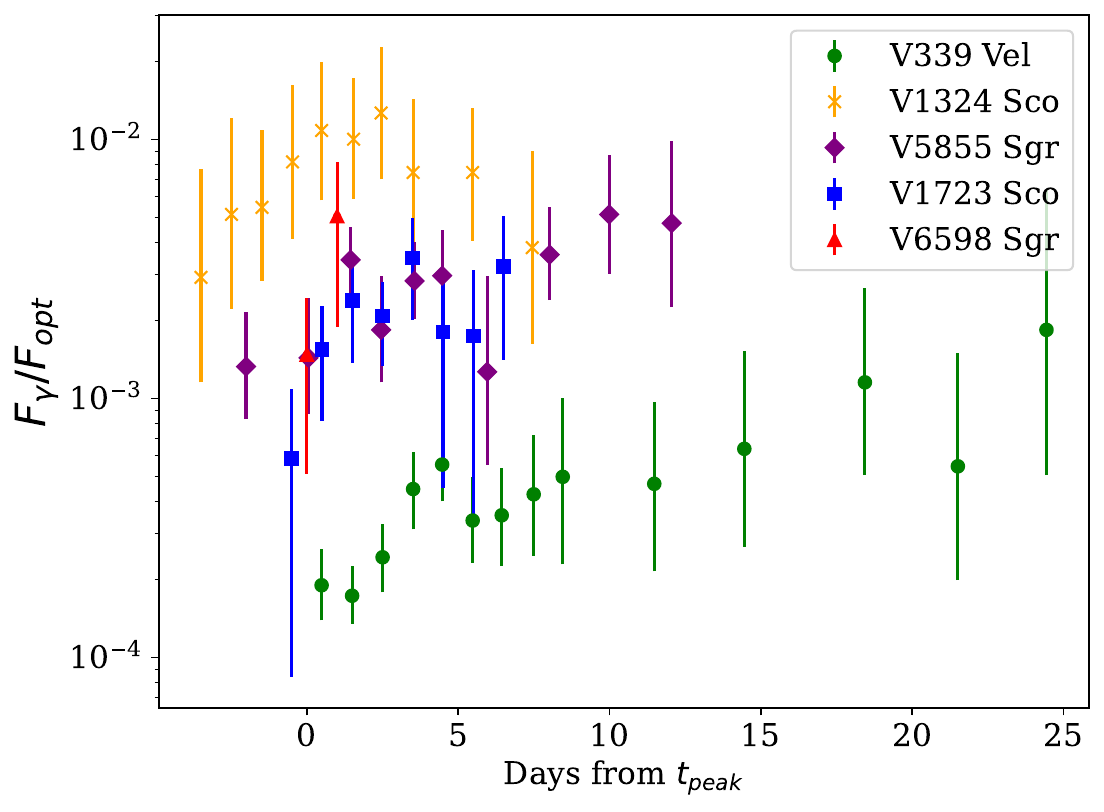}
    \caption{\label{fig:ratio} {$F_{\gamma}/F_{\text{opt}}$ ratio for several classical novae. $t_{peak}$ is the time when the peak in optical data is reached. The ratios from V1723 Sco and V6598 Sgr are calculated using the $\gamma$-ray fluxes and V-band magnitudes from Figure \ref{fig:lightcurves}. The optical fluxes are extinction corrected. For both novae, the uncertainties are at 1$\sigma$ level and are dominated by the uncertainties on the $\gamma$-ray flux. The other ones are taken from the Supplementary Figure 3 from \citet{lietal2017}. }}
\end{figure}

We estimate the $F_{\gamma}/F_{\text{opt}}$ ratio for both novae. In the case of V1723 Sco, the ratio can be estimated using 8 successive days (from $t_0 +1 $ to $t_0 + 9$, a period with significant Fermi flux points and available V-band data) of detections in both $\gamma$-ray and optical data. The optical flux $F_{\text{opt}}$ is averaged over each 1-day time bin used for the Fermi analysis using the V-band magnitude from Figure \ref{fig:lightcurves}. That leads to an average ratio during the 8-days period of $10^{-2.7}$ that varies between 10$^{-3.2}$ and $10^{-2.5}$ (see Figure \ref{fig:ratio}). For V6598 Sgr, we can only estimate the $F_{\gamma}/F_{\text{opt}}$ ratio during the two days with $\gamma$-ray detections. We obtain a mean value of $10^{-2.6}$. 
We compare the $F_{\gamma}/F_{\text{opt}}$ ratio with other classical $\gamma$-ray novae, taken from \citet{lietal2017}, in Figure \ref{fig:ratio}. The parameter $t_{peak}$ is the time when the maximum magnitude is reached in optical and is offset by two days relative to $t_{disc}$ for V1723 Sco , while there is no significant delay for V6598 Sgr. Both V1723 Sco and V6598 Sgr are comparable with the three other novae, with an average ratio of 10$^{-3}$ and an increase of the ratio over the first days of the emission.

\subsection{X-ray data}\label{X-ray data}
The layers of matter where the shocks are created are known to be opaque to soft X-rays \citep{lietal2017}. This is in agreement, in both cases, with their detection 2 weeks after the optical outburst \citep{2024ATel16444....1S,2023ATel16172....1N}. However, the hard X-rays (from thermal or non-thermal emission) could pass through the layer and could be seen by NuSTAR. V1723 Sco has been observed for 63 ks, between Feb 10.299 ($t_{disc} + 1.472$) and Feb 11.767 ($t_{disc} + 2.94$) by \citet{2024ATel16444....1S} that resulted in a non detection with an upper limit on the unabsorbed flux in the energy range of 3-78 keV of $6\times 10^{-14}$ erg cm$^{-2}$ s$^{-1}$. One can estimate the monochromatic flux at 20 keV, $\nu F_{\nu}(20\text{ keV)}$ assuming a power-law with an index of 2 for $dN/dE$. That leads to an upper limit of the monochromatic flux of $\nu F_{\nu} (20 \text{ keV)} \lesssim 2\times 10 ^{-14}$ erg cm$^{-2}$ s$^{-1}$. During the same time period, the monochromatic flux at 100 MeV measured with the Fermi-LAT can also be estimated using a PowerLaw2 (fit of the flux but not the amplitude) for a point source at the nova position. The PowerLaw2 model is preferred over the PLExpCutoff  because there is not enough statistics during this short time interval to fit the cutoff energy. The monochromatic flux at 100 MeV is so $\nu F_{\nu} (100 \text{ MeV})\approx 2\times 10 ^{-10}$ erg cm$^{-2}$ s$^{-1}$. That leads to a ratio $F_X/F_{\gamma} \lesssim 10^{-4}$. This value will be discussed in the following part.

\begin{figure}[H]
     \includegraphics[width=0.9\columnwidth]{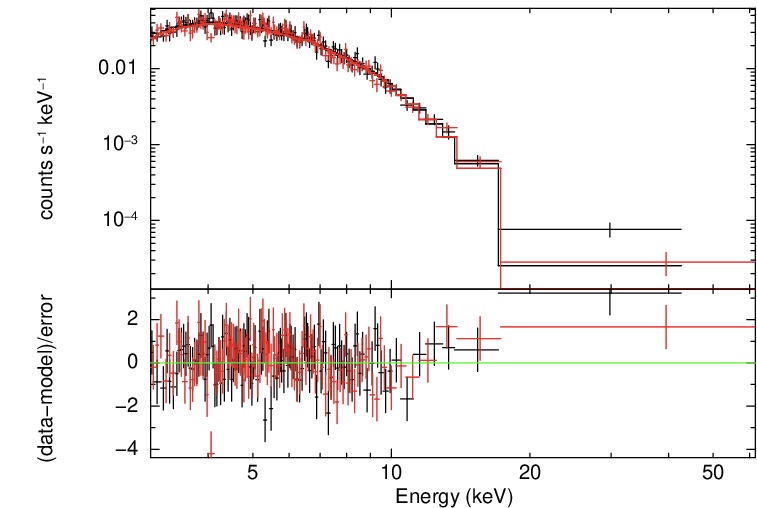}
    \caption{\label{fig:nustar} {NuSTAR spectrum with best model (see text) of the nova V1723 Sco 47 days after the outburst. The data collected with the focal plane modules A and B are plotted in black and red, respectively.}}
\end{figure}

The second NuSTAR observation of V1723 Sco started on 2024 Mar 26.729 ($t_{disc} + 46.902$) and lasted for 22 ks. It revealed bright hard X-ray emission 47 days after the nova outburst. A heavily-absorbed thermal emission with $k_bT = 3.7 \pm 0.1$ keV with a non-solar abundance fit the data well (Figure \ref{fig:nustar}). The abundances were taken from the analysis of V906 Car \citep{2020MNRAS.497.2569S}, a bright nova also detected by the LAT \citep{2018ATel11546....1J}. These non-solar abundances can be explained by the mixing of heavy material from the WD into the accreted matter shell, which presumably has near-solar abundances. We derive an integrated flux, between 3 and 78 keV, of $1.7 \pm 0.1 \times 10^{-11}$erg cm$^{-2}$ s$^{-1}$, which is one of the brightest flux values ever found for a classical nova in the hard X-ray band. No non-thermal emission counterpart was found and the spectrum is similar to the one from other classical novae in the early phase of the outburst \citep{2019ApJ...872...86N,2020MNRAS.497.2569S,2022MNRAS.514.2239S,Her2021...Sokolovsky}. This detection was not coincident in time with the 3 sigma detection 55 days after the outburst with the LAT, but suggests that hot shocked matter could still be present even months after the outburst.

\section{Discussion}
\subsection{Nature of the particles accelerated at the shock \label{chap.3.1}}

During the outburst of a nova, several shocks are created and charged particles, mostly protons and electrons, can be accelerated via DSA. These accelerated charged particles can then interact with the ambient medium and radiate at several wavelengths. The main $\gamma$-ray emission processes are the decay of $\pi^0$ produced by $p-p$ interaction, Bremsstrahlung and Inverse Compton scattering of electrons.
The $\gamma$-ray data alone do not allow discriminating that emission scenario is dominating the $\gamma$-ray flux. However, \citet{metzger_gamma-ray_2015} studied cases of a hadronic and leptonic scenario and introduced a general condition based on the $L_{\gamma}/L_{\text{opt}}$ ratio and, with the hypothesis of a radiative shock characterized by a minimum non-thermal particle acceleration efficiency, find
\begin{equation}
\epsilon_{\text{nth}}> \epsilon_{\text{nth,min}} = \frac{1}{\epsilon_{\gamma}}\frac{L_{\gamma}}{L_{\text{opt}}},
\end{equation}
with $\epsilon_{\text{nth}}$ the fraction of shock power used to accelerate non-thermal particles and $\epsilon_{\gamma}$ the fraction of the power of non-thermal particle emitted in the $\gamma$-ray band. For both hadronic and leptonic scenarios, we can assume $\epsilon_{\gamma}$ $\lesssim$ 0.2 \citep{metzger_gamma-ray_2015} and so, using the ratio $F_{\gamma}/F_{\text{opt}}$ calculated in Section \ref{chap.lgamma/lopt}, we obtain $\epsilon_{\text{nth}}$ $\gtrsim$ 10$^{-2}$ for the two novae. This value is coherent with a hadronic scenario, which can reach about $\epsilon_{\text{nth}} \sim 0.1$ in non-relativistic shocks, comparable to those in supernovae remnants (example of Tycho in \citet{morlino_strong_2012}). However, it is too high for the leptonic scenario. Indeed, the ratio of the energy placed into relativistic protons to that in relativistic electrons is estimated, in astrophysical shocks like SNRs, to be of the order of $10^{-4} - 10^{-2}$ which leads to a $\epsilon_{\text{nth}}$ for electrons of about $10^{-5}-10^{-3}$. So, in the hypothesis of a radiative shock, the hadronic scenario is preferred to explain the $\gamma$-ray emission. In the case of V1723 Sco, the ratio $F_X/F_{\gamma}\approx 10^{-4}$ also supports this hypothesis \citep{vurmandmetzger}. This low ratio disfavors a leptonic model.

\subsection{Constraining the physical properties of V1723 Sco using multi-wavelength data}
For V1723 Sco, the $\gamma$-ray emission is sufficiently bright to constrain the parameters of an emission model through a fit to the $\gamma$-ray data. In contrast, this was not feasible for V6598 Sgr due to the insufficient number of significant $\gamma$-ray flux points. The SED from Figure \ref{fig:SEDs} was fitted with a Pion Decay emission model implemented in the python module \texttt{naima} developed by \citet{naima}. 
The Pion Decay model assumes a spectral distribution of protons, a density of target $n_H$ for the $p-p$ interaction and a distance $d$ from the observer (taken from §\ref{distance}). The optical spectroscopy of the nova on days 2024 Feb 11.368 and Feb 12.362 ($t_0$ + 2.368 and $t_0$ + 3.362) by \citet{2024ATel16454....1H} reveal an expanding shell velocity $v_{\text{sh}}$ of 3200 $\pm$ 200 and 1800 $\pm$ 100 km s$^{-1}$ respectively. Since the Fermi-LAT data is integrated over a 2-week time period, $v_{\text{sh}}$ is approximated to $v_{\text{sh}}\sim$2$\times$10$^{3}$ km s$^{-1}$. The radius of the expanding shell is consequently defined as  
\begin{equation}
    R_{\text{sh}} = v_{\text{sh}}t \approx 4\times 10^{13} \ \text{cm},
    \label{eq:radius}
\end{equation}
approximately 2 days  after the start of the outburst. The density of target $n_H$ in this condition can be approximated by the density of the ejecta, $n_{\text{sh}}$, in an expanding shell of thickness $dR = R/4$ \citep{metzger_gamma-ray_2015}
\begin{equation}
        n_H=n_{\text{sh}} \approx \frac{M_{ej}}{\pi R_{\text{sh}}^{3}f_{\Omega}\mu m_p},
        \label{eq:density}
\end{equation}  
with $m_p$ the proton mass, $M_{\text{ej}}$ the total mass ejected, $\mu$ the mean molecular weight of a fully ionized gas for the environment in classical nova \citep{2020MNRAS.497.2569S}, and $f_{\Omega}$ the fraction of total solid angle of the ejecta \citep{metzger_gamma-ray_2015}. The value of $dR$ is not well constrained, and various values have been used in the literature (e.g., $dR/R = 0.2$ in \citet{2018ApJ...853...27M} and $dR/R = 0.75$ in \citet{2004A&A...420..571C}). 

Assuming a constant expansion speed, one can calculate the mean density during the 15-day time window of the $\gamma$-ray emission
\begin{multline}
    <n_{\text{sh}}> = \frac{1}{15}\int^{15}_{0}n_{\text{sh}}dt \\
    \approx  7 \times 10^{11} \left(\frac{M_{ej}}{10^{-4}M_{\odot}}\right) \left(\frac{f_{\Omega}}{0.5} \right)^{-1} \left (\frac{\mu}{0.74} \right)^{-1} \text{cm}^{-3}.
\end{multline}
This constant velocity, when used in the shell density calculation, assuming a mass of $M_{ej} \approx 10^{-4} M_{\odot}$, provides column density of $N_H = 2 \times 10^{22}$ cm$^{-2}$ that agree with the one measured 33 days and 111 days after the discovery \citep{2024ATel16641....1L}.

\begin{figure}[H]
     \includegraphics[width=1\columnwidth]{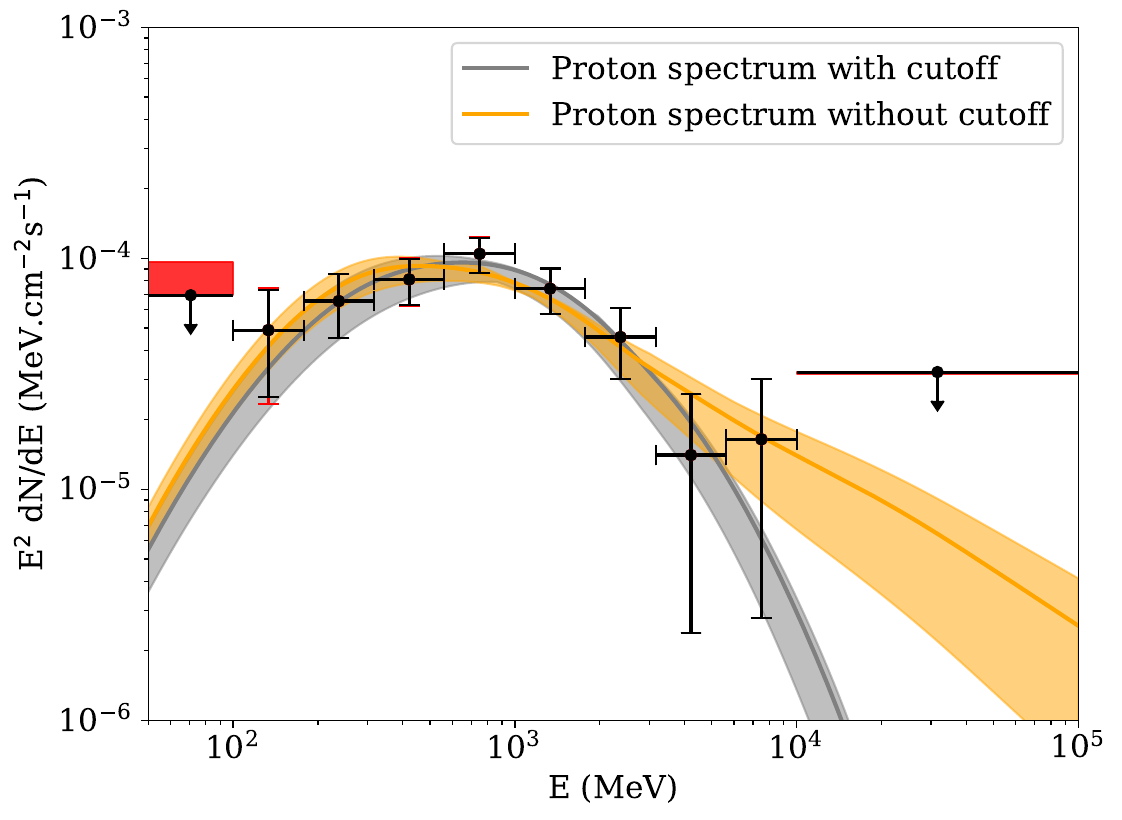}
    \caption{\label{fig:naima_PD} {Fits of the V1723 Sco $\gamma$-ray emission using a hadronic model from \texttt{naima}. Flux points are taken from Figure \ref{fig:SEDs}. The gray and yellow areas show the 1$\sigma$ uncertainty on the posterior distribution for a PLExpCutoff and PL protons spectrum respectively (see text).}}
\end{figure}

For the spectral distribution of injected protons, a PL and a PLExpCutoff model were tested. One can see that the curvature in the $\gamma$-ray spectrum at $E<500$ MeV can be reproduced by the PionDecay model assuming a simple PL spectrum for the accelerated protons (yellow curve in Figure \ref{fig:naima_PD}), due to the decrease of the $p-p$ interaction cross-section at low energy. However, the value of the resulting proton index ($s_p$) is $2.9 \pm 0.2$, which is very unlikely if we assume that protons are accelerated via DSA. Additionally, we expect a cutoff energy in the proton spectrum due to the limited size of the accelerator. This value can be estimated using the Hillas criterion, by replacing the radius of the system $R_{\text{sh}}$ by the size of the layer ionized by the free-free emission which can allow magnetic field amplification and protons acceleration \citep{metzger_gamma-ray_2015}. Using values presented above for V1723 Sco, the cutoff energy in the proton spectrum, $E_{\text{max}}$, is estimated to $E_{\text{max}}\sim 10$ GeV if we assume a high value ($10^{-2}-10^{-4}$) of the fraction of the magnetic pressure compared to the thermal pressure of the shock (2.3.1 of \citet{metzger_gamma-ray_2015}). During the fit, the amplitude and the slope of the spectral distribution of protons are free to evolve. To determine the best fit parameters, \texttt{naima} used the python module \texttt{emcee} \citep{emcee} which is a Python implementation of Goodman $\&$ Weare’s Affine Invariant Markov chain Monte Carlo (MCMC) Ensemble sampler. The uncertainties on the parameters are estimated with the first and third quartile of the posterior distribution. Results of the fit are shown in Figure \ref{fig:naima_PD}. 

The best fit parameters for the PLExpCutoff are $s_p=1.8 \pm 0.3$ and an amplitude of 10$^{33.1\pm0.3}$ eV$^{-1}$. That leads to a total energy of the proton distribution above 1.2 GeV of 
\begin{equation}
    W_p = 4.4\pm3.7 \times 10^{39} \left(\frac{M_{ej}}{10^{-4}M_{\odot}}\right) ^{-1} \left(\frac{f_{\Omega}}{0.5}\right)  \left(\frac{\mu}{0.74}\right) \left(\frac{d}{1.9 \text{ kpc}}\right)^{2} \text{ erg.}
\end{equation}

The uncertainties on $W_p$ arise from the spectral parameters of the particle distribution and do not account for the uncertainties on the distance and the target density. The statistics do not allow observing an evolution of the parameters during the two weeks if we consider the same environment of the nova during the entire time period. $W_p$ can be compared to the value of the total energy available in the shock. Assuming a fully radiative shock, one can find an upper limit for the shock energy $W_{\text{sh}}\lesssim W_{opt} \approx 10^{43} $ erg during the 15 days of Fermi analysis. That leads to a fraction of $\gtrsim $ 0.1 $\%$ of the total energy of the shock used to accelerate protons. It is one order of magnitude lower than the estimation using \citet{metzger_gamma-ray_2015} in §\ref{chap.3.1}. That could be explained by the estimation of the target density, $n_H$, which is not necessarily distributed homogeneously in the expanding shell. Additionally, the $\gamma$-ray absorption was also neglected (see §\ref{chap.abs_gam} for discussion). And even in the case of a radiative shock, a large fraction of the optical emission might come from thermal emission of the unshocked nova ejecta. A more refined model could yield the expected value of the $W_p/W_{\text{sh}}$ ratio.

\subsection{$\gamma$-ray absorption}\label{chap.abs_gam}
\begin{figure}[H]
     \includegraphics[width=1.0\columnwidth]{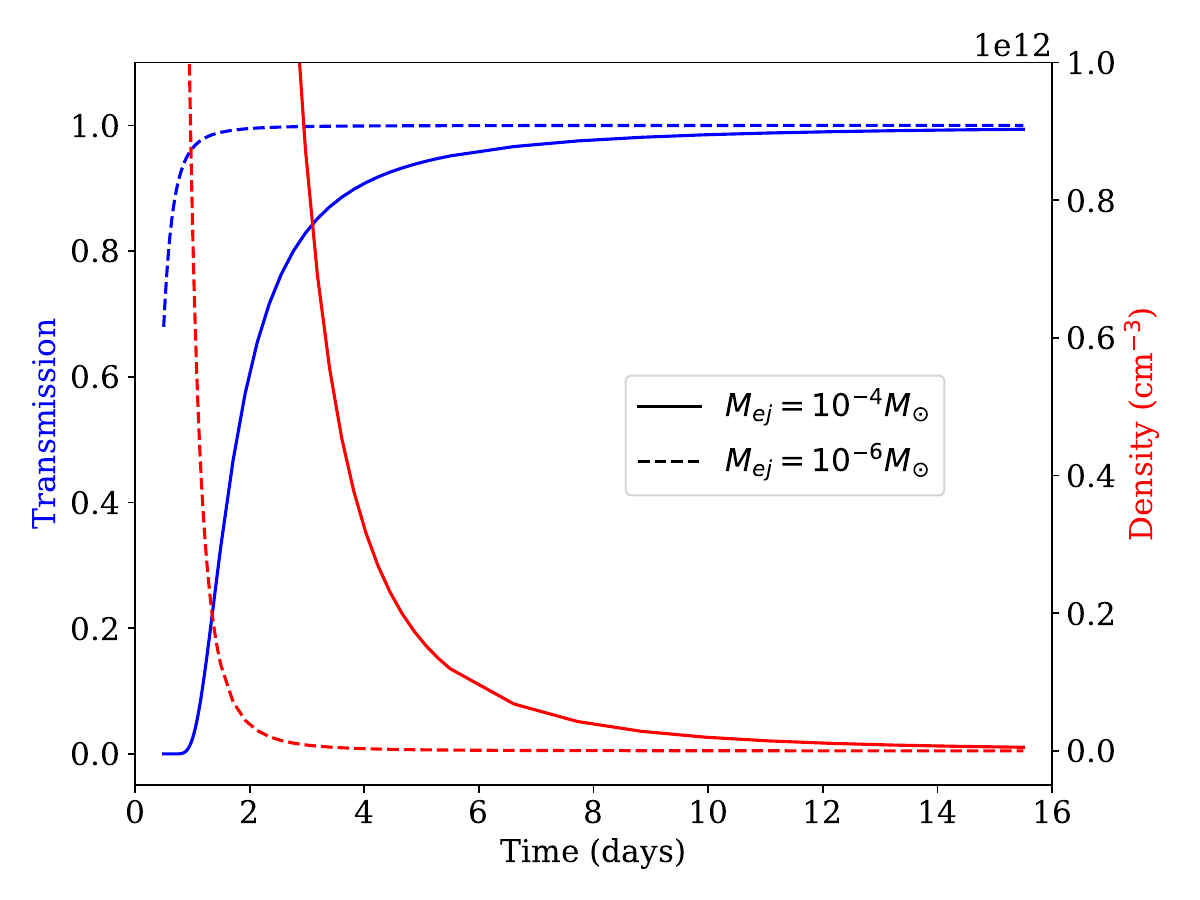}
    \caption{\label{fig:transmission} {Target density and transmission for two different $M_{ej}$ and $\Delta t$ = 0.5 days. Transmission is due to Compton scattering and pair creation and is calculated for a $\gamma$-ray energy of 1 GeV.}}
\end{figure}

In the previous section, the absorption of $\gamma$-rays was neglected. However, at the expected target densities, $\gamma$-rays could be absorbed via Compton scattering and pair production. In order to quantify this absorption a simple model for its computation was developed. For the pair production, only the interaction with a nucleus was taken into account because the interaction with photons from the pseudo-photosphere (photon with E $\sim$ 1 eV) is only relevant for $\gamma$-ray with E > 300 GeV. The time dependent shell evolution, with a thickness $dR=R/4$, was computed to study the density profile. In this case, the interesting part is when the $\gamma$-rays are already produced and can interact with the shell. Consequently, we define the variable $\Delta t$ as the time between the beginning of the expansion and the start of $\gamma$-ray interaction with matter. The ejected mass is also not well constrained and we assumed a typical value of 10$^{-4} M_{\odot}$ in the previous section. However, this value can vary between 10$^{-6}$ and 10$^{-3} M_{\odot}$ \citep{chomiuk_new_2021} and significantly influence the resulting $\gamma$-ray emission. Consequently, $M_{ej}$ was a parameter of our model. Finally, the velocity of the expanding shell was modeled by an exponential increase $v_{\text{sh}}(t) = v_{\text{sh,Max}} (1-\exp{(t/\tau)})$ with $v_{\text{sh,Max}} = 2 \times 10^3 $ km s$^{-1}$ and $\tau = 1$ day. One can now estimate the absorption using an exponential law for the transmission. The results at 1 GeV are presented in Figure \ref{fig:transmission}.

As expected, the absorption is negligible for low $M_{ej}$ but can have an influence during the first few days, particularly for the previously used value of 10$^{-4} M_{\odot}$. However, this is true only during the initial days of the emission (only 20$\%$ of the $\gamma$-ray emission is absorbed by day 4). Unfortunately, the multi-wavelength analysis does not provide a better constraint on the ejected mass, and it is not possible to constrain it using Fermi data alone. However, we add this absorption calculation to our previous analysis and we find that the total proton energy $W_p$ increases by a factor of $\sim 4$ assuming the same ejecta mass of 10$^{-4} M_{\odot}$. This is in agreement with our previous conclusion of the under-estimation of the $W_p/W_{\text{sh}}$ ratio.

This is a very simple model that does not account for the spatial distribution of $\gamma$-ray production, nor for the orientation and true shape of the ejecta. Providing a more precise treatment would require introducing several new hypotheses, which cannot be reliably constrained, and that is why we do not address it in further detail.

\subsection{Delayed detection of V1723 Sco with NuSTAR and Fermi}

Following up on the potential evidence for a late $\gamma$-ray signal in V1723 Sco, we study its potential implications here. The index of 2.1 from the Fermi analysis suggests that the system could accelerate high-energy particles even 50 days after the nova eruption. The NuSTAR detection a few days earlier could indicate the presence of hot matter heated by a new shock, which might be responsible for accelerating these particles. The NuSTAR data analysis suggests an intrinsic column density of $N_H = 2.1 \pm 0.1 \times 10^{21}$ cm$^{-2}$ on top of the adopted galactic line of sight absorption of $8.48   \times 10^{21}$ cm$^{-2}$ derived from 21 cm radio data of \citet{2005A&A...440..775K}. If we assume that the ejecta still expand at $v_{\text{sh,Max}}$ \citep{2019MNRAS.483.3773S} even 50 days after the eruption, the density of target calculated as before, will be only around 10$^{7}$ cm$^{-3}$ in the expanding shell. This leads to a total proton energy required to produce the $\gamma$-ray emission seen by Fermi which exceeds the value we calculated above. That means that the accelerated particles have to be accelerated elsewhere. A better target could be the matter accumulated from the hot WD wind, which is not completely diluted, given that the SSS phase started only 50 days after these two detections \citep{2024ATel16641....1L}. However, it is complicated to define the geometry of this accumulated matter in order to have a target density and a more complete set of multi-wavelength data is needed.

\subsection{Strong magnetic field in V6598 Sgr}\label{chap.magnetic_field}
There is not enough Fermi-LAT data to constrain the spectrum of the accelerated particles in V6598 Sgr. However, this very brief $\gamma$-ray emission could be interestingly linked to the nature of the binary system. To date, only 2 other novae detected by Fermi have been identified as Intermediate Polars (IPs): V1674 Her and V407 Lup. In contrast to V6698 Sgr, the IP classification was derived from optical periodicity analyses. The brightest, V1674 Her (Her 2021), was detected for only 18 hours, starting 6 hours after the start of the eruption \citep{Her2021...Sokolovsky}. The spectral shape of the $\gamma$-ray emission of the nova was not peculiar, but it was the shortest $\gamma$-ray emission ever observed for a classical nova. The second case, V407 Lup, was reported in \citet{V407Lup..Gordon} at a detection level of 3$\sigma$. 
The analysis of the 3 days of V407 Lup's $\gamma$-ray emission was done following the procedure presented in this work, no particular shape was found and a simple power-law reproduced the Fermi-LAT data well. With their brief 3-day durations, these three novae represent the shortest $\gamma$-ray emitting novae ever observed by Fermi-LAT.

The magnetic field of IP can exceed $10^6$ G, which could lead to rapid cooling of electrons but also protons via synchrotron losses following 
\begin{equation}
        t_{cool,sync}^{e} \approx 4 \times 10^{-7} \left (\frac{B}{10^6 \text{ G}}\right ) ^{-2} \left ( \frac{E}{1\text{ GeV}} \right )^{-1} \text{ s}
\end{equation} 
and
\begin{equation}
            t_{cool,sync}^{p} \approx \left ( \frac{m_p}{m_e}\right )^{5/2} t_{cool,sync}^{e} \approx 60 \text{ s},
\end{equation}
with  $t_{cool,sync}^{e}$ and $t_{cool,sync}^{p}$ the cooling time via synchrotron losses for electrons and protons respectively. However, this value of $B$ is estimated at the surface of the WD and is proportional to $R^{-3}$. And, following the explanation described for V1723 Sco, at $R = 10^{13}$ cm, the magnetic field from the WD in the magnetic dipole approximation is only few $\mu$G.

V6598 Sgr has also the particularity to have no $\gamma$-ray emission below 1 GeV (see Figure \ref{fig:SEDs}) and that could be explained by a high density of gas in the region of the accelerated particles, which could lead to an opacity in the $\gamma$-ray range (see §\ref{chap.abs_gam}). This difference with other classical nova, such as V1723 Sco, could indicate another location of the particle acceleration site.

\section{Conclusion}

Since the discovery of classical novae as $\gamma$-ray sources, the hypothesis of accelerated protons via diffusive shock acceleration at the interface between a slow ejecta and a fast wind from the hot shocked white dwarf dominates. The recent observation of the nova V1723 Sco 2024 supports this hypothesis. The complete analysis of the Fermi-LAT data has shown a significant detection during two weeks in $\gamma$-rays well-fitted by a hadronic scenario in the nova environment. This emission is coincident with the optical data reported by the AAVSO. Despite its short duration, the nova V6598 Sgr presented a faint $\gamma$-ray emission during the optical peak. However, like the other novae, the non detection of non-thermal hard X-ray emission is still a challenging question. Moreover, the indication of late time emission by NuSTAR and Fermi-LAT (not simultaneously) more than 40 days after the eruption for the nova V1723 Sco shows that acceleration of charged particles to high energy could last more than a month (or start again), and, this time, with the expected hard X-ray component. Finally, the peculiar case of novae with an intermediate polar, which appeared to have a very short emission in optical, and the shortest emissions detected by the Fermi-LAT, can not be explained by assuming a strong magnetic field and hints at a different acceleration region.
\\

\begin{acknowledgements} 
The \textit{Fermi} LAT Collaboration acknowledges generous ongoing support
from a number of agencies and institutes that have supported both the
development and the operation of the LAT as well as scientific data analysis.
These include the National Aeronautics and Space Administration and the
Department of Energy in the United States, the Commissariat \`a l'Énergie Atomique
and the Centre National de la Recherche Scientifique / Institut National de Physique
Nucl\'eaire et de Physique des Particules in France, the Agenzia Spaziale Italiana
and the Istituto Nazionale di Fisica Nucleare in Italy, the Ministry of Education,
Culture, Sports, Science and Technology (MEXT), High Energy Accelerator Research
Organization (KEK) and Japan Aerospace Exploration Agency (JAXA) in Japan, and
the K.~A.~Wallenberg Foundation, the Swedish Research Council and the
Swedish National Space Board in Sweden.
 
Additional support for science analysis during the operations phase is gratefully
acknowledged from the Istituto Nazionale di Astrofisica in Italy and the Centre
National d'\'Etudes Spatiales in France. This work performed in part under DOE
Contract DE-AC02-76SF00515.

Research at the Naval Research Laboratory is supported by NASA DPR S-15633-Y.

KS is supported by NASA award 80NSSC24K1168.

LC acknowledges support from NASA grants 80NSSC23K0497, 80NSSC23K1247, and 80NSSC25K7334.

J.L.S. acknowledges support from NASA awards 80NSSC25K7068 and 80NSSC25K7082.

\end{acknowledgements}

\bibliography{biblio}{}
\bibliographystyle{aa}

\end{document}